\documentclass[a4paper]{spie}
\usepackage{graphicx}
\usepackage{amsfonts,amssymb,amsbsy,amsmath}

\usepackage{graphicx}

\usepackage{bbm}

\usepackage[normalem]{ulem}

\usepackage{color}

\title{Chiral-stress-energy-momentum tensor for covariant description of spin and torque densities of light} 
\author{Mikko Partanen$^1$ and Jukka Tulkki$^{2}$\skiplinehalf
$^1$Photonics Group, 
Department of Electronics and Nanoengineering, Aalto 
University,\\ P.O. Box 13500, 00076 Aalto, Finland\skiplinehalf
$^2$Engineered Nanosystems Group, School of Science, Aalto 
University,\\ P.O. Box 12200, 00076 Aalto, Finland\\
}

\begin{document}

\maketitle

\begin{abstract}
The measurement of the spin angular momentum of circularly polarized light by Beth [Phys. Rev. 50, 115 (1936)] can be explained by using a microscopic torque density. However, the experiment does not resolve the space- and time-dependent evolution of the spin density of light and the wave plate and the covariant form of the microscopic torque density. Here we focus on the covariant description of the helicity, spin, and torque densities of light in materials using the chiral-stress-energy-momentum tensor. We also perform simulations of Gaussian light pulses in quarter-wave-plate geometries made of birefringent and dielectric materials.
\end{abstract}

\section{Introduction}

The spin angular momentum of circularly polarized light was quantitatively measured in a seminal work by Beth in 1936 \cite{Beth1936}.
Beth proved experimentally that the torque on a quarter-wave plate when linearly polarized thermal light from a tungsten ribbon filament incomes to the quarter-wave plate and leaves it circularly polarized corresponds to the gain of spin angular momentum of light equal to $\pm\hbar$ per photon. Beth also presented a theoretical model reproducing this experimental result by assuming that the spin-torque density on the homogeneous birefringent material of the quarter-wave plate is given by \cite{Beth1936}
\begin{equation}
 \boldsymbol{\tau}_\mathrm{e}=\mathbf{P}\times\mathbf{E}.
 \label{eq:torquedensity0}
\end{equation}
Here $\mathbf{P}$ is the polarization field and $\mathbf{E}$ is the electric field. In the experiment, by observing the rotation of the quarter-wave plate, Beth could only determine the volume integral of the spin-torque density. The space- and time-dependent evolution of the spin-torque density inside the birefringent material was not resolved. The boundary conditions at the fringe of the quarter-wave plate were not discussed in detail in Beth's experiment. The behavior of the fields at the boundary has been shown to have an important contribution to the conservation of the total angular momentum in Beth's experiment by Allen and Padgett \cite{Allen2002} in their response to the question ``Does plane wave not carry a spin?" by Khrapko \cite{Khrapko2001}.

In phenomenological terms, we associate the spin density of light to the property of light to give to a material local classical spin angular momentum whenever, in a scattering process, a small differential part of the light wave is absorbed or converted to another polarization state. We neglect the atomic structure of the material and mean by the spin angular momentum of the microscopic material body, its rotation around its center of mass. To highlight the point-rotational nature of the torque density in Eq.~\eqref{eq:torquedensity0}, we call it the spin-torque density in distinction to the classical torque density $(\mathbf{r}-\bar{\mathbf{r}})\times\mathbf{f}$, which we call the Poynting torque density, and which is determined by the force density $\mathbf{f}$ and the point of reference $\bar{\mathbf{r}}$. However, note that, at the microscopic level, the origin of all force and torque densities is on the Lorentz forces on elementary charges as discussed in Sec.~\ref{sec:microscopic} below. In this sense, both the Poynting torque density and the spin-torque density are quantities that arise in the effective field theory, where the description of the microscopic structure of the material is replaced by the effective polarization and magnetization fields that are spatially smooth functions over the atomic or molecular structure. Consequently, in this work, we assume that the fields are spatially averaged over the atomic or molecular structure, but not over the wavelength of light. This is a broadly used assumption in classical optics.

In this work, we develop a theory of classical spin of light in materials that applies to a general time-varying field, and not only to the special case of the monochromatic field, which has been widely studied in previous literature \cite{Landau1982,Sakurai1967,Dirac1958,Messiah1961,Alpeggiani2018,Bliokh2017b}. We study the possibility that, following the general principle of measurability in classical physics, one could measure the spin-torque density of the material as a function of the position and time. This would, in principle, enable determining the exact position- and time-dependent form of the spin-torque density and the position- and time-dependent spin density of classical light inside birefringent materials, linear dielectrics, and vacuum. For the case of the monochromatic field, we also formulate the covariant description of the angular momentum transfer by introducing the chiral-stress-energy-momentum (CSEM) tensor.

\section{Microscopic derivation of the torque density}
\label{sec:microscopic}

Theoretical determination of the spin-torque density in Eq.~\eqref{eq:torquedensity0} is relatively straightforward. We consider the spin torque on an atomic dipole as a sum of Lorentz-force-generated torques on the positive and negative charges of the dipole with respect to the center of the dipole. The force applied on a single electric charge $\pm q_\mathrm{e}$ in an electric field $\mathbf{E}$ and magnetic flux density $\mathbf{B}$ at position $\mathbf{r}_{\pm q_\mathrm{e}}$ and time $t$ is known as the Lorentz force, and it is given by
\begin{equation}
 \mathbf{F}_{\pm q_\mathrm{e}}(t)=\pm q_\mathrm{e}\Big[\mathbf{E}(\mathbf{r}_{\pm q_\mathrm{e}},t)+\frac{d\mathbf{r}_{\pm q_\mathrm{e}}}{dt}\times\mathbf{B}(\mathbf{r}_{\pm q_\mathrm{e}},t)\Big].
 \label{eq:Lorentz}
\end{equation}
We approximate the external electric field $\mathbf{E}(\mathbf{r}_{\pm q_\mathrm{e}},t)$ by its value at the center $\mathbf{r}_0$ of the electric dipole made of two charges as $\mathbf{E}(\mathbf{r}_{\pm q_\mathrm{e}},t)=\mathbf{E}(\mathbf{r}_0,t)$. The torque on a single charge is given by $\boldsymbol{\Gamma}_{\pm q_\mathrm{e}}(t)=\mathbf{r}_{\pm q_\mathrm{e}}\times\mathbf{F}_{\pm q_\mathrm{e}}(t)$. Noting that the second term of Eq.~\eqref{eq:Lorentz} gives a negligible contribution to the torque on an electric dipole in its rest frame, the net torque on the two charges of the dipole is then given by
\begin{align}
 \boldsymbol{\Gamma}_\mathrm{e}(t) &=\boldsymbol{\Gamma}_{+q_\mathrm{e}}(t)+\boldsymbol{\Gamma}_{-q_\mathrm{e}}(t)\nonumber\\
 &=q_\mathrm{e}\mathbf{r}_{+q_\mathrm{e}}\times\mathbf{E}(\mathbf{r}_0,t)
 -q_\mathrm{e}\mathbf{r}_{-q_\mathrm{e}}\times\mathbf{E}(\mathbf{r}_0,t)\nonumber\\
 &=\mathbf{p}\times\mathbf{E}(\mathbf{r}_0,t).
 \label{eq:dipoletorque}
\end{align}
Here we have defined the dipole moment $\mathbf{p}$ of the dipole as $\mathbf{p}=q_\mathrm{e}(\mathbf{r}_{+q_\mathrm{e}}-\mathbf{r}_{-q_\mathrm{e}})$. Multiplying both sides of Eq.~\eqref{eq:dipoletorque} by the number density of the dipoles, denoted by $n_\mathrm{a}$, defining the spin-torque density as $\boldsymbol{\tau}_\mathrm{e}=n_\mathrm{a}\boldsymbol{\Gamma}_\mathrm{e}$ and the polarization field by $\mathbf{P}=n_\mathrm{a}\mathbf{p}$, we obtain the equation of the spin-torque density in Eq.~\eqref{eq:torquedensity0}. Note that, in the case of nonbirefringent, nondispersive linear materials, the orientation of the induced dipole moment follows the orientation of the electric field. Consequently, the cross product on the last line of Eq.~\eqref{eq:dipoletorque} is zero. Therefore, the spin-torque density is zero for nonbirefringent linear materials.

\section{Spin density of light}

We assume that the fields in the material are initially zero. Consequently, the free and bound charge and current densities in the material are also initially zero. Then, an electromagnetic wave is assumed to enter the material. Regarding the angular momentum density of classical light, we generalize the results of light in vacuum by Cameron \emph{et al.}~\cite{Cameron2012} to apply to more general linear and birefringent materials. We propose that the space- and time-dependent helicity density $\rho_{\mathfrak{h},\mathrm{em}}$, spin density $\boldsymbol{\rho}_{\mathbf{S},\mathrm{em}}$, and the spin-stress tensor $\mathbf{n}_{\mathbf{S},\mathrm{em}}$, called ij-infra-zilches by Cameron \emph{et al.}~\cite{Cameron2012}, are given in the laboratory frame by
\begin{equation}
 \rho_{\mathfrak{h},\mathrm{em}}=\frac{1}{2c}(\mathbf{A}\cdot\mathbf{H}-\mathbf{C}\cdot\mathbf{E}),
 \label{eq:uS}
\end{equation}
\begin{equation}
 \boldsymbol{\rho}_{\mathbf{S},\mathrm{em}}=\frac{1}{2}(\mathbf{D}\times\mathbf{A}+\mathbf{B}\times\mathbf{C}),
 \label{eq:spindensity}
\end{equation}
\begin{equation}
 \mathbf{n}_{\mathbf{S},\mathrm{em}}
 =\!\frac{1}{2}\Big[(\mathbf{A}\cdot\mathbf{H}-\mathbf{C}\cdot\mathbf{E})\mathbf{I}-\mathbf{A}\otimes\mathbf{H}-\mathbf{H}\otimes\mathbf{A}+\mathbf{C}\otimes\mathbf{E}+\mathbf{E}\otimes\mathbf{C}\Big].
 \label{eq:spinstresstensor}
\end{equation}
The constitutive relations of the electric and magnetic field strengths $\mathbf{E}$ and $\mathbf{H}$, electric and magnetic flux densities $\mathbf{D}$ and $\mathbf{B}$, and the polarization and magnetization fields $\mathbf{P}$ and $\mathbf{M}$ are given in the laboratory frame by $\mathbf{D}=\varepsilon_0\mathbf{E}+\mathbf{P}$ and $\mathbf{B}=\mu_0\mathbf{H}+\mu_0\mathbf{M}$, where $\varepsilon_0$ and $\mu_0$ are the permittivity and permeability of vacuum. Furthermore, in a nondispersive material, the electric flux density $\mathbf{D}$ and the magnetic field strength $\mathbf{H}$ can be written in terms of the fields $\mathbf{E}$ and $\mathbf{B}$ by the conventional linear relations $\mathbf{D}=\varepsilon_0\boldsymbol{\varepsilon}_\mathrm{r}\mathbf{E}$ and $\mathbf{H}=(\mu_0\boldsymbol{\mu}_\mathrm{r})^{-1}\mathbf{B}$, where $\boldsymbol{\varepsilon}_\mathrm{r}$ and $\boldsymbol{\mu}_\mathrm{r}$ are the relative permittivity and permeability tensors.

Under our assumption that the electric and magnetic fields in the material are initially zero at time $t=-\infty$, the fields $\mathbf{A}$ and $\mathbf{C}$ in Eqs.~\eqref{eq:uS}--\eqref{eq:spinstresstensor} are unambiguously determined by the electric and magnetic fields as $\mathbf{A}=-\int_{-\infty}^t\mathbf{E}dt'$ and $\mathbf{C}=-\int_{-\infty}^t\mathbf{H}dt'$. Thus, $\mathbf{A}$ and $\mathbf{C}$ \emph{have no gauge freedom associated to them}. It follows that $\mathbf{A}$ and $\mathbf{C}$ are related to the electric and magnetic fields by the derivative relations $\mathbf{E}=-\frac{\partial}{\partial t}\mathbf{A}$ and $\mathbf{H}=-\frac{\partial}{\partial t}\mathbf{C}$. From Faraday's law, we find that $\mathbf{B}=\nabla\times\mathbf{A}$, and from the Ampere-Maxwell law, in the absence of free electric current density, we find that $\mathbf{D}=-\nabla\times\mathbf{C}$. Thus, $\mathbf{A}$ and $\mathbf{C}$ can be identified with the well-known vector potential \cite{Jackson1999} and the dual vector potential \cite{Bliokh2013b}, which are here effectively forced to be in the radiation gauge.

\section{\label{sec:CSEMconservation}Conservation laws of helicity and spin}

Using the spin-related quantities in Eqs.~\eqref{eq:uS}--\eqref{eq:spinstresstensor}, we obtain the dynamical equations of the helicity and spin angular momentum densities of the electromagnetic field, given by
\begin{equation}
 \frac{1}{c}\frac{\partial\rho_{\mathfrak{h},\mathrm{em}}}{\partial t}+\nabla\cdot\boldsymbol{\rho}_{\mathbf{S},\mathrm{em}}=-\frac{\varphi}{c^2},
 \label{eq:helicitydensityconservationlawfield}
\end{equation}
\begin{equation}
 \frac{\partial\boldsymbol{\rho}_{\mathbf{S},\mathrm{em}}}{\partial t}+\nabla\cdot\mathbf{n}_{\mathbf{S},\mathrm{em}}=-\boldsymbol{\tau}.
 \label{eq:torquedensityconservationlawfield}
\end{equation}
Here $\varphi$ is the helicity-conversion density and $\boldsymbol{\tau}$ is the spin-torque density, which is the generalization of $\boldsymbol{\tau}_\mathrm{e}$, given in Eq.~\eqref{eq:torquedensity0}. Equation \eqref{eq:helicitydensityconservationlawfield} is the differential form of the conservation law of helicity and Eq.~\eqref{eq:torquedensityconservationlawfield} is the differential form of the conservation law of the spin angular momentum. The conserved quantities, called constants of motion, are the volume integrals of the corresponding densities of the full system of the field and the material. The corresponding conservation laws have been studied in the case of the electromagnetic field in vacuum in the absence of the source terms by Cameron \emph{et al.}~\cite{Cameron2012}.

% Taking the Fourier transformation of Eq.~\eqref{eq:helicitydensityconservationlawfield} in the absence of the source term gives $-\frac{1}{c}(-i\omega)\rho_{\mathfrak{h},\mathrm{em}}-i\mathbf{k}\cdot\boldsymbol{\rho}_{\mathbf{S},\mathrm{em}}=0$, where $\omega$ is the angular frequency and $\mathbf{k}$ is the wave vector. Solving this equation for $\rho_{\mathfrak{h},\mathrm{em}}$ shows that our definition of the helicity density for a plane wave corresponds to the conventional definition of the helicity density as the component of the spin density parallel to the wave vector, given by $\rho_{\mathfrak{h},\mathrm{em}}=\boldsymbol{\rho}_{\mathbf{S},\mathrm{em}}\cdot\mathbf{k}/k_0$, where $k_0=\omega/c$ \cite{Landau1982,Sakurai1967,Dirac1958,Messiah1961,Alpeggiani2018}.

Using the conservation laws in Eqs.~\eqref{eq:helicitydensityconservationlawfield} and \eqref{eq:torquedensityconservationlawfield}, the definitions of $\rho_{\mathfrak{h},\mathrm{em}}$, $\boldsymbol{\rho}_{\mathbf{S},\mathrm{em}}$, and $\mathbf{n}_\mathbf{S}$ in Eqs.~\eqref{eq:uS}--\eqref{eq:spinstresstensor}, the definitions of the vector potentials $\mathbf{A}$ and $\mathbf{C}$, the radiation gauge relations $\nabla\cdot\mathbf{A}=0$ and $\nabla\cdot\mathbf{C}=0$, and the constitutive relations of the fields, we straightforwardly obtain the helicity-conversion and spin-torque densities, given by
\begin{equation}
 \varphi=\frac{1}{2}\Big(\mathbf{J}_\mathrm{m}\cdot\mathbf{A}-\frac{\mathbf{J}_\mathrm{e}\cdot\mathbf{C}}{\varepsilon_0}\Big),
 \label{eq:varphi}
\end{equation}
% \begin{align}
%  \boldsymbol{\tau} &=\mathbf{P}\times\mathbf{E}+\mathbf{M}\times\mathbf{B}
%  +\frac{1}{2}\Big[(\nabla\cdot\mathbf{H})\mathbf{A}+(\nabla\cdot\mathbf{A})\mathbf{H}\nonumber\\
%  &\hspace{0.4cm}-(\nabla\cdot\mathbf{E})\mathbf{C}-(\nabla\cdot\mathbf{C})\mathbf{E}\Big].
%  \label{eq:torquedensity5}
% \end{align}
\begin{equation}
 \boldsymbol{\tau}=\mathbf{P}\times\mathbf{E}+\mathbf{M}\times\mathbf{B}.
 \label{eq:torquedensity5}
\end{equation}
Here $\mathbf{J}_\mathrm{e}=\frac{\partial\mathbf{P}}{\partial t}+\nabla\times\mathbf{M}$ is the bound electric current density and $\mathbf{J}_\mathrm{m}=\frac{\partial\mathbf{M}}{\partial t}-c^2\nabla\times\mathbf{P}$ is the bound magnetic current density. Equation \eqref{eq:torquedensity5} of the spin-torque density is the generalization of Eq.~\eqref{eq:torquedensity0} for a general instantaneous electromagnetic field in materials that are both polarizable and magnetizable.

Since the physical quantities in Eqs.~\eqref{eq:helicitydensityconservationlawfield} and \eqref{eq:torquedensityconservationlawfield} are determined by the same electric and magnetic fields as the electromagnetic energy and momentum densities, the helicity and spin conservation laws are not independent of the energy and momentum conservation laws. In this sense, helicity and spin are quantities that are conserved simultaneously with energy and momentum. The helicity and spin are needed for determining the local spin-torque density experienced by the material through Eq.~\eqref{eq:torquedensityconservationlawfield}, which is different from the orbital torque density $(\mathbf{r}-\bar{\mathbf{r}})\times\mathbf{f}$, determined by the force density $\mathbf{f}$ and the point of reference $\bar{\mathbf{r}}$.

% \textcolor{red}{\uline{Illustration of the last terms of the spin-torque density for a Gaussian beam.}}

\section{Spin and torque equivalences}

\subsection{Spin equivalence}

For Gaussian light pulses in vacuum, it is well known that the volume integral of the spin density of the field over the full light pulse is equal to the volume integral of the Poynting angular momentum density $(\mathbf{r}-\mathbf{r}_\mathrm{ce})\times\mathbf{G}_\mathrm{em}$, where $\mathbf{G}_\mathrm{em}=\mathbf{E}\times\mathbf{H}/c^2$ is the momentum density of the field and $\mathbf{r}_\mathrm{ce}$ is the center of energy of the pulse \cite{Partanen2018a}. This equivalence is written as
\begin{equation}
 \int\boldsymbol{\rho}_{\mathbf{S},\mathrm{em}}d^3r=\int(\mathbf{r}-\mathbf{r}_\mathrm{ce})\times\mathbf{G}_\mathrm{em}d^3r.
 \label{eq:spin-equivalence}
\end{equation}
It can be noted that the momentum density of the field can also be written as $\mathbf{G}_\mathrm{em}=\boldsymbol{\rho}_{\mathbf{p},\mathrm{em}}+\frac{1}{2}\nabla\times\boldsymbol{\rho}_{\mathbf{S},\mathrm{em}}$. The first term of this expression is the orbital momentum density and the second term is the spin momentum density \cite{Berry2009,Ohanian1986,Mita2000,Wittig2009,Bliokh2010,Bliokh2013b,Bliokh2017b}. For Gaussian light pulses, only the spin momentum density contributes to the integral on the right of Eq.~\eqref{eq:spin-equivalence}. In studies of the total angular momentum density of the coupled field-material state of light in materials, the Poynting angular momentum density is associated with the angular momentum density of the atomic mass density wave driven forward by the optical force density \cite{Partanen2018a}.

\subsection{Torque equivalence}

At all instants of time, the net torque on an arbitrarily-shaped material body in an electromagnetic field is equal to the volume integral of the spin-torque density $\boldsymbol{\tau}$, which is equal to the volume integral of the classical Poynting torque density $(\mathbf{r}-\bar{\mathbf{r}})\times\mathbf{f}$, where $\mathbf{f}$ is the force density given in Appendix \ref{apx:SEM}, as
\begin{equation}
 \int_V\boldsymbol{\tau}d^3r=\int_V(\mathbf{r}-\bar{\mathbf{r}})\times\mathbf{f}d^3r.
 \label{eq:torque-equivalence}
\end{equation}
The choice of the reference point $\bar{\mathbf{r}}$ is arbitrary. The choice of $\bar{\mathbf{r}}$ does not, however, contribute to the value of the integral over the full volume of the material. The dependence of $(\mathbf{r}-\bar{\mathbf{r}})\times\mathbf{f}$ on $\bar{\mathbf{r}}$ indicates that this classical Poynting torque density has no physical meaning as a local quantity. In contrast, the spin-torque density $\boldsymbol{\tau}$ has a clear physical meaning as a local quantity, which is also indicated by the microscopic derivation in Sec.~\ref{sec:microscopic}.

In the case of a piecewise homogeneous material, one can divide the volume $V$ into interior volume $V_\mathrm{i}$ and the boundary volume $V_\mathrm{b}$. The integral of $\boldsymbol{\tau}$ over the infinitesimal boundary volume is zero, while the integral of $(\mathbf{r}-\bar{\mathbf{r}})\times\mathbf{f}$ over the boundary volume is generally nonzero and can be converted into a surface integral expressed in terms of the electromagnetic stress-tensor $\boldsymbol{\mathfrak{T}}_\mathrm{em}$, given in Appendix \ref{apx:SEM}, as
\begin{equation}
 \int_V(\mathbf{r}-\bar{\mathbf{r}})\times\mathbf{f}d^3r=\int_{V_\mathrm{i}}(\mathbf{r}-\bar{\mathbf{r}})\times\mathbf{f}d^3r-\int_{\partial V}(\mathbf{r}-\bar{\mathbf{r}})\times\boldsymbol{\mathfrak{T}}_\mathrm{em}\cdot d\mathbf{A}.
 \label{eq:torque-division}
\end{equation}
The values of the two integrals on the right of Eq.~\eqref{eq:torque-division} separately depend on the choice of the origin even though the sum of these terms is independent of the origin. This clearly demonstrates that $(\mathbf{r}-\bar{\mathbf{r}})\times\mathbf{f}$ has no local physical meaning as discussed above.

\section{\label{sec:CSEM}Chiral-stress-energy-momentum tensor of light}

Next, we define the CSEM tensor for a monochromatic field. The CSEM tensor is given in units of the energy density in the same way as the conventional stress-energy-momentum (SEM) tensor, given in Appendix \ref{apx:SEM}. However, the chiral energy density, chiral momentum density, and chiral stress density components of the CSEM tensor are not related to the true energy density, momentum density, and stress tensor, which are components of the SEM tensor.

The quantities $\rho_{\mathfrak{h},\mathrm{em}}$, $\boldsymbol{\rho}_{\mathbf{S},\mathrm{em}}$, and $\mathbf{n}_{\mathbf{S},\mathrm{em}}$ in Eqs.~\eqref{eq:uS}--\eqref{eq:spinstresstensor} do not form a covariant second-rank tensor on their own. We propose that the CSEM tensor of light with angular frequency $\omega$ is given by
\begin{align}
 (\widetilde{T}_\mathrm{em})^{\mu\nu}
=\left[\!\begin{array}{cccc}
 \omega\rho_{\mathfrak{h},\mathrm{em}} & \omega\boldsymbol{\rho}_{\mathbf{S},\mathrm{em}}\\
 \omega\boldsymbol{\rho}_{\mathbf{S},\mathrm{em}} & (\omega/c)\mathbf{n}_{\mathbf{S},\mathrm{em}}
\end{array}\!\right]\!.
 \label{eq:CSEMem}
\end{align}
The CSEM tensor of light in Eq.~\eqref{eq:CSEMem} is found to enable the covariant description of the spin angular momentum and torque densities of light. In the general inertial frame, the components $\rho_{\mathfrak{h},\mathrm{em}}$, $\boldsymbol{\rho}_{\mathbf{S},\mathrm{em}}$, and $\mathbf{n}_{\mathbf{S},\mathrm{em}}$ obtain dependencies on the velocity field of the material, which are not given in Eqs.~\eqref{eq:uS}--\eqref{eq:spinstresstensor}. We leave these dependencies as a topic of further work.

\section{Simulation of electromagnetic fields in a birefringent material}

We investigate the propagation of light in a birefringent material, where the relative permittivity and permeability tensors are generically written as
\begin{equation}
 \boldsymbol{\varepsilon}_\mathrm{r}=
 \left(\begin{array}{ccc}
 \varepsilon_{xx} & \varepsilon_{xy} & \varepsilon_{xz}\\
 \varepsilon_{yx} & \varepsilon_{yy} & \varepsilon_{yz}\\
 \varepsilon_{zx} & \varepsilon_{zy} & \varepsilon_{zz}
 \end{array}\right),
 \hspace{0.4cm}
 \boldsymbol{\mu}_\mathrm{r}=
 \left(\begin{array}{ccc}
 \mu_{xx} & \mu_{xy} & \mu_{xz}\\
 \mu_{yx} & \mu_{yy} & \mu_{yz}\\
 \mu_{zx} & \mu_{zy} & \mu_{zz}
 \end{array}\right).
\end{equation}
The elements of these tensors are assumed to be real valued. The typical case when the relative permittivity tensor is a diagonal matrix $\boldsymbol{\varepsilon}_\mathrm{r}=\mathrm{diag}(\varepsilon_{xx},\varepsilon_{yy},\varepsilon_{zz})$ and the relative permeability tensor is equal to an identity matrix, given by $\boldsymbol{\mu}_\mathrm{r}=\mathrm{diag}(1,1,1)$, is a special case of the more general solution. Furthermore, for a quarter-wave plate, in which light propagates parallel to the $z$ axis and whose fast axis is aligned parallel to the $x$ axis, we have $\varepsilon_{xx}=n_\mathrm{fast}^2$ and $\varepsilon_{yy}=n_\mathrm{slow}^2$. For $\varepsilon_{zz}$, we use $\varepsilon_{zz}=n_\mathrm{ave}^2$, where $n_\mathrm{ave}=\frac{1}{2}(n_\mathrm{fast}+n_\mathrm{slow})$.

In all simulations of the present work, the wavelength of light is assumed to be $\lambda_0=1064$ nm. The refractive indices of the fast and slow axes of the quarter-wave plate are $n_\mathrm{fast}=1.5$ and $n_\mathrm{slow}=1.6$. The thickness of the birefringent material in the quarter-wave plate is $d_\mathrm{slab}=\lambda_0/[4(n_\mathrm{slow}-n_\mathrm{fast})]=2.66$ $\mu$m. The antireflective coating layers at the entry and exit interfaces of the quarter-wave plate have a refractive index $n_\mathrm{coat}=1.2248$ and thickness $d_\mathrm{coat}=\lambda_0/(4n_\mathrm{coat})=217.17$ nm. The harmonic average of the Poynting vector magnitude at the beam or pulse center is $\langle S_\mathrm{0}\rangle=1.2496\times 10^{11}W/m^2$. In the laser pulse studies, the relative standard deviation of the central angular frequency $\omega_0$ is set to $\Delta\omega_0/\omega_0=0.05$.

\begin{figure}[b]
\centering
\includegraphics[width=0.55\textwidth]{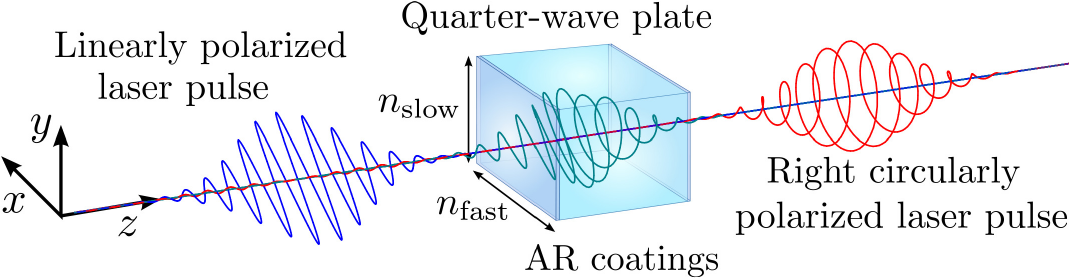}
\caption{\label{fig:geometry}
Schematic illustration of the electric field of a linearly polarized Gaussian light pulse propagating through a quarter-wave plate, in which the pulse becomes right circularly polarized. The wave plate has antireflective (AR) coatings to minimize reflections. The electric field curves are taken from the simulation corresponding to Fig.~\ref{fig:1D}.}
\end{figure}

\subsection{Propagation of a plane-wave Gaussian pulse through a quarter-wave plate}

We search for a transverse electromagnetic plane wave solution propagating parallel to the $z$ axis in a birefringent material and whose electric field amplitude is made of two linear polarization components parallel to the $x$ and $y$ axes. The electric field amplitude is given by
\begin{equation}
 \mathbf{E}(z,t) =\mathrm{Re}[i\omega\mathcal{A}(e^{i(k_1z-\omega t)}\hat{\mathbf{x}}+e^{i(k_2z-\omega t)}\hat{\mathbf{y}})].
\end{equation}
Here $\omega$ is the angular frequency, $\mathcal{A}$ is a normalization constant, $k_1$ and $k_2$ are the wavenumbers of the two components, and $\hat{\mathbf{x}}$ and $\hat{\mathbf{y}}$ are unit vectors parallel to the $x$ and $y$ axes. The refractive indices $n_1$ and $n_2$ of the two components of the wave are determined by the wavenumbers as $n_1=k_1/k_0$ and $n_2=k_2/k_0$, where $k_0=\omega/c$ is the wavenumber in vacuum, in which $c$ is the speed of light.

A propagating plane wave solution to Maxwell's equations in the absence of free or bound electric and magnetic charges is found when the elements of the relative permittivity and permeability tensors are not completely arbitrary, but they satisfy the constraints, given by
\begin{equation}
 \varepsilon_{zx} =\varepsilon_{zy}=\mu_{zx} =\mu_{zy}=0,\hspace{1.0cm}
 \mu_{xy} =\frac{\varepsilon_{yx}\mu_{xx}}{\varepsilon_{xx}},
 \hspace{1.0cm}
 \mu_{yx} =\frac{\varepsilon_{xy}\mu_{yy}}{\varepsilon_{yy}}.
\end{equation}
Then, the refractive indices $n_1$ and $n_2$ of the two components of the wave and the average refractive index $n_\mathrm{ave}$, corresponding to the average velocity of the wave in the birefringent material, are given by
\begin{equation}
 n_1=\sqrt{\varepsilon_{xx}\mu_{yy}-\varepsilon_{yx}\mu_{yx}},\hspace{1.0cm}
 n_2=\sqrt{\varepsilon_{yy}\mu_{xx}-\varepsilon_{xy}\mu_{xy}},\hspace{1.0cm}
 n_\mathrm{ave}=\frac{n_1+n_2}{2}.
\end{equation}
For $\boldsymbol{\mu}_\mathrm{r}=\mathrm{diag}(1,1,1)$, we thus obtain $\varepsilon_{xx}=n_1^2$ and $\varepsilon_{yy}=n_2^2$. Furthermore, as discussed above, we define $n_1=n_\mathrm{fast}$ and $n_2=n_\mathrm{slow}$, in which case $\boldsymbol{\varepsilon}_\mathrm{r}=\mathrm{diag}(n_\mathrm{fast}^2,n_\mathrm{slow}^2,n_\mathrm{ave}^2)$.

As an example, we simulate the propagation of a plane-wave Gaussian laser pulse through a quarter-wave plate coated with antireflective coatings as illustrated in Fig.~\ref{fig:geometry}. The incident Gaussian pulse is assumed to be linearly polarized along $\frac{1}{2}(\hat{\mathbf{x}}+\hat{\mathbf{y}})$. In the quarter-wave plate, the linear polarization is transformed into right-handed circular polarization. In the right-handed circular polarization, the electric field vector at a fixed position rotates clockwise as a function of time when viewed from the point of view of the source.

\begin{figure}[b]
\centering
\includegraphics[width=0.75\textwidth]{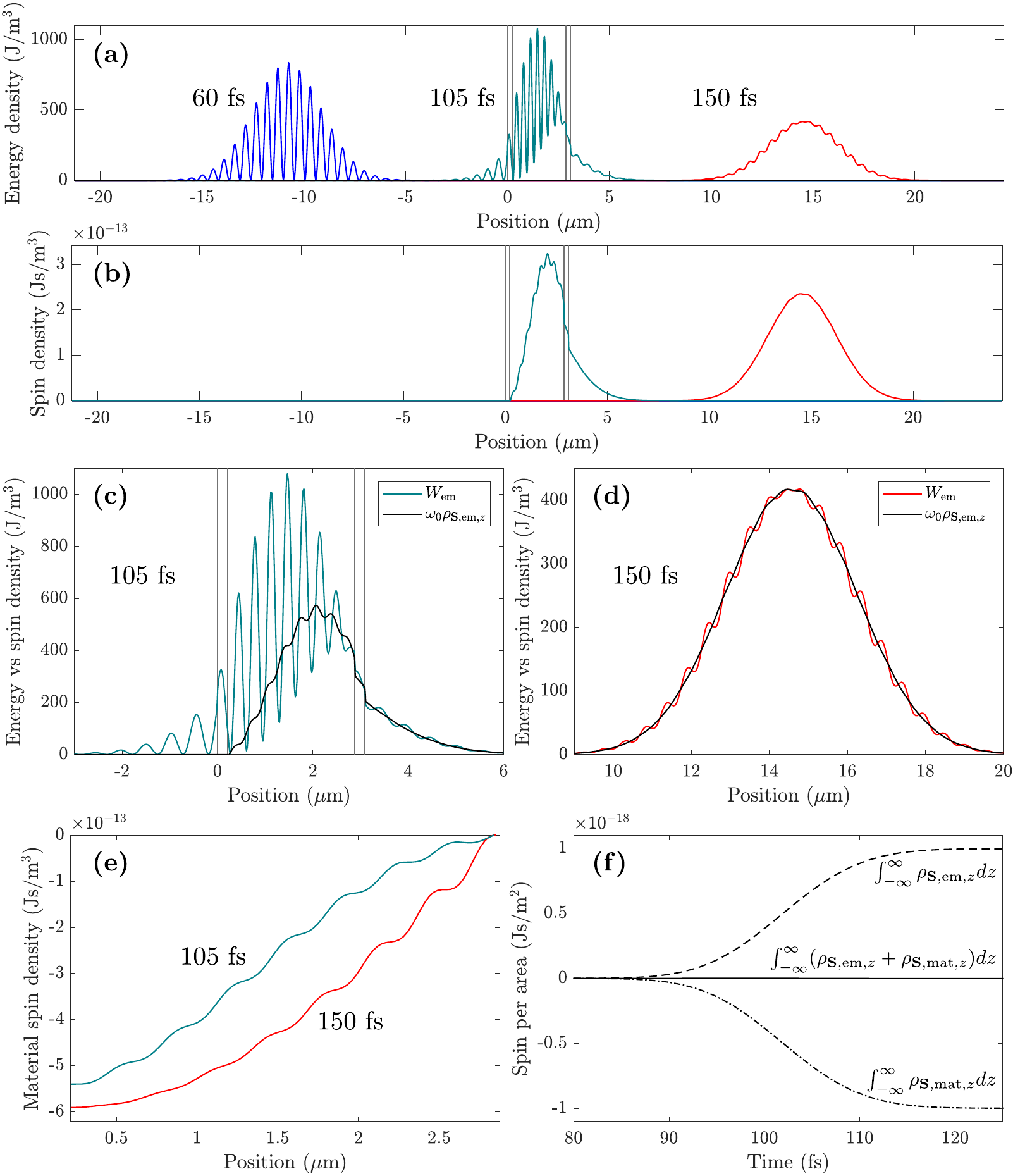}
\caption{\label{fig:1D}
(a) Energy density and (b) spin density $z$ component of a temporally Gaussian plane-wave laser pulse propagating to the right at three instants of time. (c) Energy density and scaled spin density $z$ component at 105 fs and (d) the same quantities at 150 fs. (e) The distributions of the material spin density due to the spin-torque density of the Gaussian pulse at 105 fs and 150 fs. (f) Time dependencies of the electromagnetic and material spin densities integrated over the $z$ axis. The vertical solid lines indicate the positions of the material interfaces.}
\end{figure}

Figure \ref{fig:1D}(a) shows the instantaneous energy density of the Gaussian pulse at three instants of time. At 60 fs, the pulse propagates toward the quarter-wave plate in vacuum. At 105 fs, the center of the pulse is inside the quarter-wave plate. At 150 fs, the pulse has fully passed through the wave plate and it is propagating away from it in vacuum. Due to the antireflective coatings, only a negligible part of the pulse has reflected from the quarter-wave plate. This part cannot be seen in the scale of the figure. Before the wave plate, the energy density of the pulse has variations under the Gaussian envelope proportional to the square of a sinusoidal function. After the wave plate, the energy density of the pulse is almost Gaussian, and it has only small variations under the Gaussian envelope. These variations originate from the different propagation velocities of the Gaussian envelopes of the two linear polarization components in the quarter-wave plate.

Figure \ref{fig:1D}(b) shows the instantaneous spin density of the pulse at the three instants of time corresponding to the energy density in Eq.~\ref{fig:1D}(a). Before the quarter-wave plate, the pulse is linearly polarized, and therefore, the spin density is zero. Inside the quarter-wave plate, the pulse is transformed so that it becomes right-handed circularly polarized and the spin density increases. Figures \ref{fig:1D}(c) and \ref{fig:1D}(d) compare the energy and spin densities of the pulse inside and after the quarter-wave plate. The spin density is scaled by the centaral angular frequency $\omega_0$. At the exit interface of the quarter-wave plate in Fig.~\ref{fig:1D}(c) and after the wave plate in Fig.~\ref{fig:1D}(d), the difference between the energy and spin densities is small.

Figure \ref{fig:1D}(e) depicts the spin density of the quarter-wave plate material resulting from the torque density that the birefringent material experiences due to the Gaussian pulse. The spin density of the material is shown at 105 fs and at 150 fs. The values at 150 fs are smaller since, at this instant of time, the pulse has fully passed the wave plate. Positive spin is given to the field and negative spin is given to the material. The total spin of the system is zero throughout the simulation. The spin densities of the electromagnetic field, the material, and their sum integrated over the $z$ axis are shown as a function of time in Fig.~\ref{fig:1D}(f).

\subsection{Propagation of a 3D Gaussian beam through a quarter-wave plate}

Next, we consider a 3D Gaussian beam with a finite beam waist radius propagating through the quarter-wave plate with the focal plane located at the the entry interface of the wave plate at $z=0$ $\mu$m. The geometry of the quarter-wave plate is the same as that illustrated in Fig.~\ref{fig:geometry}. The beam waist radius is $w_0=3\lambda_0/2$. Since the electric and magnetic fields of the laser pulse and the material parameters of the birefringent wave-plate material have no azimuthal symmetry, full 3D simulations are necessary.

\begin{figure}
\centering
\includegraphics[width=\textwidth]{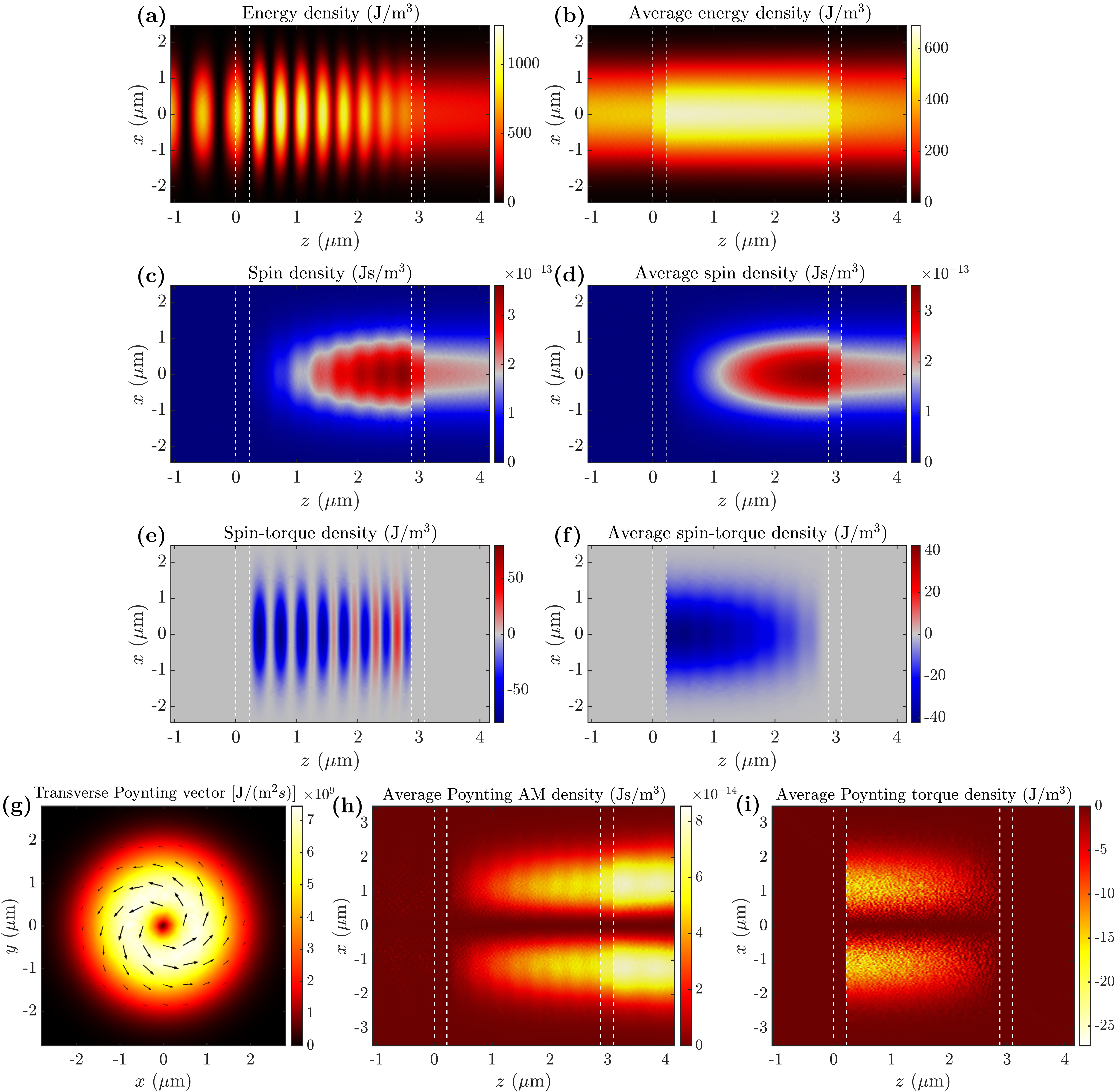}\\
\caption{\label{fig:3D}
(a) Energy density, (b) time average of the energy density, (c) spin density $z$ component, (d) time average of the spin density $z$ component, (e) spin-torque density $z$ component, (f) the time average of the spin-torque density $z$ component, (g) transverse components of the Poynting vector (arrows) and their magnitude (background), (h) classical Poynting angular momentum density $z$ component, and (i) the corresponding Poynting torque density $z$ component as a function of the position when a linearly polarized 3D Gaussian beam propagates through the quarter-wave plate in which it becomes circularly polarized. The quantities presented as a function of the $x$ and $y$ coordinates are in the plane after the quarter-wave plate. The vertical dashed lines indicate the positions of the material interfaces.}
\end{figure}

Figure Fig.~\ref{fig:3D}(a) shows the instantaneous energy density. The maxima of the instantaneous energy density before the wave plate in vacuum are equal to the maximum value of the energy density of the plane wave pulse of Fig.~\ref{fig:1D}(a) before the wave plate in vacuum. The Gaussian beam spreads when going away from the focal plane. However, this divergence cannot be seen in the scale of Fig.~\ref{fig:3D}(a). Figure \ref{fig:3D}(b) shows the energy density of Fig.~\ref{fig:3D}(a) time-averaged over the harmonic cycle. The time-averaging removes the variations of the energy density inside the harmonic cycle, which are associated with linear polarization.

Figure \ref{fig:3D}(c) shows the instantaneous spin density $z$ component of the Gaussian beam. The time-averaged spin density is presented in Fig.~\ref{fig:3D}(d). In analogy to the plane-wave pulse case in Fig.~\ref{fig:1D}(b), the spin density is zero before the quarter-wave plate in vacuum. The spin density increases in the wave plate, where the beam becomes right-handed circularly polarized. The instantaneous and time-averaged spin-torque densities are depicted in Figs.~\ref{fig:3D}(e) and \ref{fig:3D}(f). The instantaneous spin-torque density has large variations inside the harmonic cycle, where it obtains both positive and negative values. The values of the time-averaged spin-torque density are negative. The spin-torque density gives a negative spin to the material that is of equal magnitude to the positive spin obtained by the field.  

The transverse components of the Poynting vector and their magnitude at the exit interface of the quarter-wave plate are shown in Fig.~\ref{fig:3D}(g). These components are the origin of the Poynting angular momentum density in the integrand on the right of Fig.~\eqref{eq:spin-equivalence} calculated with respect to $r_\mathrm{ce}=(0,0,0)$ and whose time-averaged $z$ component is depicted in Fig.~\ref{fig:3D}(h). The time-average of the corresponding Poynting torque density $z$ component is presented in Fig.~\ref{fig:3D}(i). In agreement with Eq.~\eqref{eq:torque-equivalence}, it is found that the volume integral of the spin-torque density over the quarter-wave plate is equal to the volume integral of the Poynting torque density. Since the off-axis components of the Poynting vector of the Gaussian beam are not exactly normal to the exit interface, the surface integral term of Eq.~\eqref{eq:torque-division} has a contribution to the Poynting torque density.

\clearpage
\subsection{Scattering of a monochromatic plane wave from a quarter-wave plate cylinder}

Next, we consider the case opposite to the investigation of the Gaussian beam above, where the spatial extent of the beam was smaller than the quarter-wave plate in the $x$ and $y$ directions. In the opposite case, the spatial extent of the electromagnetic field in the $x$ and $y$ directions is larger than the dimensions of the quarter-wave plate. We study the scattering of a monochromatic linearly polarized plane wave from a quarter-wave-plate cylinder with a diameter of two vacuum wavelengths as $d=2\lambda_0$. Apart from the cylinder shape of the quarter-wave plate, the geometry of the materials is equal to that in Fig.~\ref{fig:geometry}.

\begin{figure}[b]
\centering
\includegraphics[width=\textwidth]{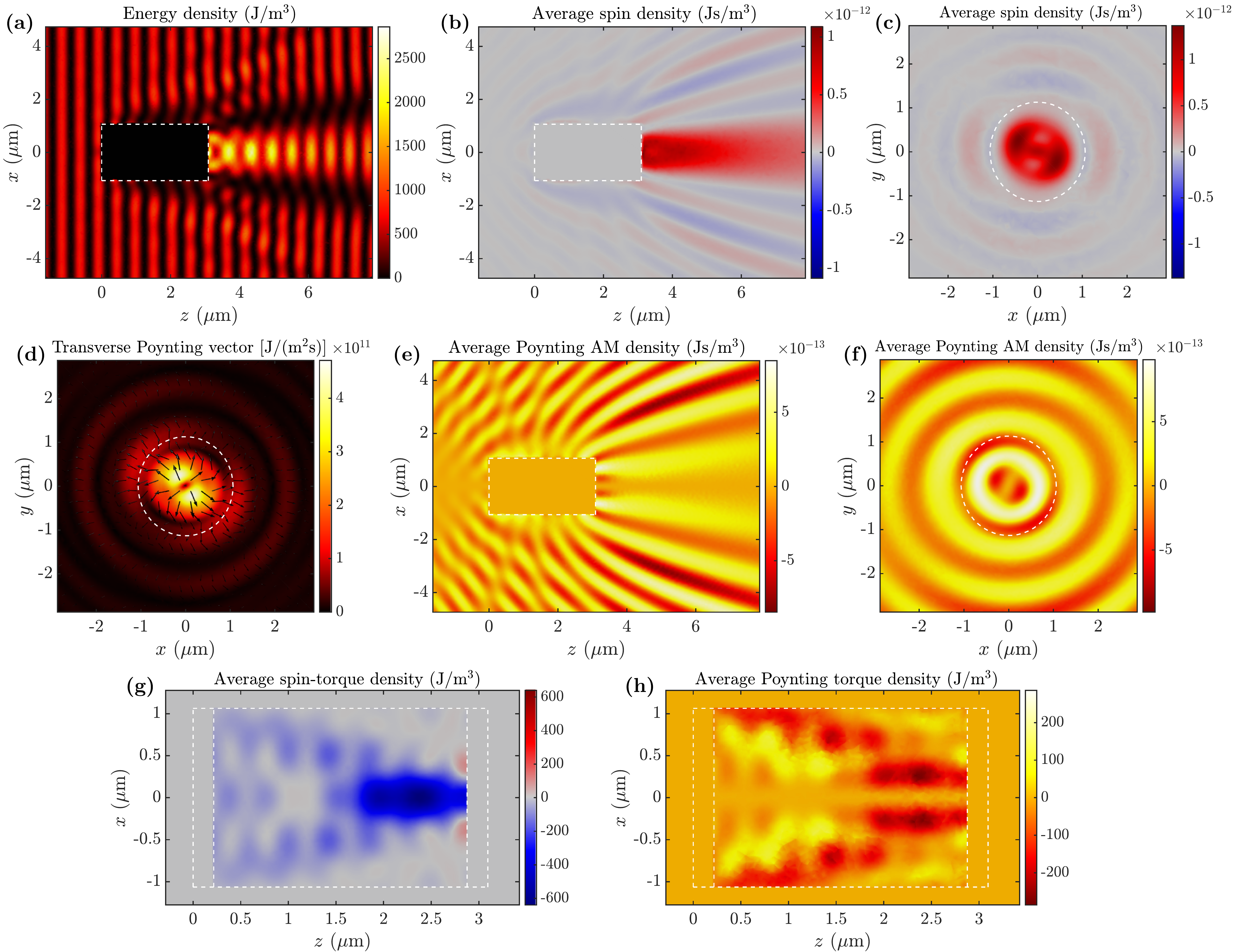}
\caption{\label{fig:scattering}
(a) Instantaneous energy density, (b) average spin density $z$ component, (c) average spin density $z$ component in the transverse plane, (d) transverse components of the Poynting vector (arrows) and their magnitude (background), (e) average Poynting angular momentum density $z$ component, (f) average Poynting angular momentum $z$ component in the transverse plane, (g) average spin-torque density $z$ component, and (h) average Poynting torque density $z$ component of a linearly polarized plane wave scattered from a quarter-wave-plate cylinder with a small diameter. The dashed rectangle and circle indicate the position of the quarter-wave-plate cylinder. The quantities presented as a function of the $x$ and $y$ coordinates are in the plane after the quarter-wave plate. The energy, spin, and Poynting angular momentum densities inside the wave plate are not shown since these quantities at certain positions inside the wave plate have values exceeding the scale of the pertinent figures.}
\end{figure}

The instantaneous energy density of the linearly polarized monochromatic field scattered from the quarter-wave-plate cylinder is shown in Fig.~\ref{fig:scattering}(a). The time-averaged spin density is presented in the $xz$ plane in Fig.~\ref{fig:scattering}(b) and in the $xy$ plane in Fig.~\ref{fig:scattering}(c). It is seen that the field that passes through the quarter-wave-plate cylinder has a notable right-handed circularly polarized contribution associated with the positive spin density. Farther from the optical axis, the field scattered from the cylinder forms regions of both positive and negative spin densities.

The transverse components of the Poynting vector at the $xy$ plane after the cylinder are shown in Fig.~\ref{fig:scattering}(d). These components are related to the formation of the Poynting angular momentum density, given by the integrand on the right of Eq.~\eqref{eq:spin-equivalence} calculated with respect to $r_\mathrm{ce}=(0,0,0)$. The $z$ component of the Poynting angular momentum density is depicted in the $xz$ plane in Fig.~\ref{fig:scattering}(e) and in the $xy$ plane in Fig.~\ref{fig:scattering}(f). Since the Poynting angular momentum density is equal to the cross-product of the position vector and the Poynting momentum density, it attenuates much slower than the transverse components of the Poynting vector in Fig.~\ref{fig:scattering}(d). The Poynting angular momentum density obtains both positive and negative values away from the axis of the cylinder, where it is zero.

The average spin-torque density $z$ component inside the quarter-wave-plate cylinder is presented in Fig.~\ref{fig:scattering}(g). The average Poynting torque density is shown in Fig.~\ref{fig:scattering}(h). While the spin-torque density is unambiguously defined, the Poynting torque density, given in the integrand on the right of Eq.~\eqref{eq:torque-equivalence}, depends on the reference point $\bar{r}$. It is associated with an interface torque density not shown in Fig.~\ref{fig:scattering}(h). In agreement with Eqs.~\eqref{eq:torque-equivalence} and \eqref{eq:torque-division}, the volume integrals of the spin and Poynting angular momentum densities over the quarter-wave-plate cylinder are found to be equal.

\section{\label{sec:conclusions}Conclusions}

In conclusion, we have studied the space- and time-dependent instantaneous and time-averaged spin and torque densities of classical light. We have performed simulations of Gaussian light pulses and beams in quarter-wave-plate geometries made of dielectric and birefringent materials. We have also presented the CSEM tensor for the covariant description of spin and torque densities of light. The conservation laws written through the four-divergence of the CSEM tensor are the helicity and spin conservation laws in a material, which are found to be consistent with the torque density used by Beth to explain his seminal measurement \cite{Beth1936} of the spin angular momentum of circularly polarized light.

\begin{acknowledgments}
This work has been funded by the Research Council of Finland under Contract No.~349971. Aalto Science-IT is acknowledged for computational resources.
\end{acknowledgments}

\appendix

\section{Stress-energy-momentum tensor of light}
\label{apx:SEM}
\setcounter{figure}{0}\renewcommand{\thefigure}{A\arabic{figure}}
\setcounter{equation}{0}\renewcommand{\theequation}{A\arabic{equation}}

The stress-energy-momentum (SEM) tensor of the electromagnetic field in a material is given by \cite{Partanen2023a}
\begin{equation}
 (T_\mathrm{em})^{\mu\nu}=\left[\begin{array}{cccc}
W_\mathrm{em} & c\mathbf{G}_\mathrm{em}\\
c\mathbf{G}_\mathrm{em} & \boldsymbol{\mathfrak{T}}_\mathrm{em}
\end{array}\right].
 \label{eq:SEMem}
\end{equation}
Here $W_\mathrm{em}$ is the electromagnetic energy density, $\mathbf{G}_\mathrm{em}$ is the electromagnetic momentum density, and $\boldsymbol{\mathfrak{T}}_\mathrm{em}$ is the electromagnetic stress tensor. In the laboratory frame, the energy density, momentum density, and the stress tensor components are given by
\begin{equation}
 W_\mathrm{em}=\frac{1}{2}(\mathbf{E}\cdot\mathbf{D}+\mathbf{H}\cdot\mathbf{B}),
 \label{eq:Wem}
\end{equation}
\begin{equation}
 \mathbf{G}_\mathrm{em}=\frac{\mathbf{E}\times\mathbf{H}}{c^2},
 \label{eq:Gem}
\end{equation}
\begin{equation}
 \boldsymbol{\mathfrak{T}}_\mathrm{em}=\frac{1}{2}\Big[(\mathbf{E}\cdot\mathbf{D}+\mathbf{H}\cdot\mathbf{B})\mathbf{I}-\mathbf{E}\otimes\mathbf{D}-\mathbf{D}\otimes\mathbf{E}-\mathbf{H}\otimes\mathbf{B}-\mathbf{B}\otimes\mathbf{H}\Big]-\frac{1}{2}(\mathbf{P}\cdot\mathbf{E}+\mathbf{M}\cdot\mathbf{B})\mathbf{I}.
 \label{eq:Tem}
\end{equation}
The energy density $W_\mathrm{em}$ is of the well-known form, the momentum density $\mathbf{G}_\mathrm{em}$ is equal to the so-called Abraham momentum density. The first term of the stress tensor $\boldsymbol{\mathfrak{T}}_\mathrm{em}$ is the conventional Abraham stress tensor and the last term of $\boldsymbol{\mathfrak{T}}_\mathrm{em}$ results from the optical electro- and magnetostriction \cite{Partanen2023a,Anghinoni2022,Anghinoni2023}.

\subsection{\label{apx:SEMconservation}Conservation laws of energy and momentum}

In terms of the components of the SEM tensor, the conservation laws of the energy and momentum of the electromagnetic field are given by \cite{Penfield1967,Partanen2023a,Partanen2021b,Kemp2017}
\begin{equation}
 \frac{1}{c^2}\frac{\partial W_\mathrm{em}}{\partial t}+\nabla\cdot\mathbf{G}_\mathrm{em}
 =-\frac{\phi}{c^2},
\end{equation}
\begin{equation}
 \frac{\partial\mathbf{G}_\mathrm{em}}{\partial t}+\nabla\cdot\boldsymbol{\mathfrak{T}}_\mathrm{em}
 =-\mathbf{f}.
 \label{eq:Gemiconservation}
\end{equation}
Here $\mathbf{f}$ is the force density and $\phi$ is the power-conversion density. These quantities are related to each other by $\phi=\mathbf{f}\cdot\mathbf{v}_\mathrm{a}$ \cite{Penfield1967}. The power-conversion and force densities can be split into the free charge and current contributions $\phi_\mathrm{L}$ and $\mathbf{f}_\mathrm{L}$ and homogeneous material contributions $\phi_\mathrm{A}$ and $\mathbf{f}_\mathrm{A}$, from which $\mathbf{f}_\mathrm{A}$ is known as the generalized Abraham force \cite{Obukhov2008,Partanen2022b}. The force density additionally has a material interface contribution $\mathbf{f}_\mathrm{int}$ and the optostriction contribution $\mathbf{f}_\mathrm{ost}$. Thus, we write the power-conversion density  $\phi$ and the force density $\mathbf{f}$ as
\begin{equation}
 \phi=\phi_\mathrm{L}+\phi_\mathrm{A},
 \hspace{0.5cm}\mathbf{f}=\mathbf{f}_\mathrm{L}+\mathbf{f}_\mathrm{A}+\mathbf{f}_\mathrm{int}+\mathbf{f}_\mathrm{ost}.
 \label{eq:phif}
\end{equation}
In terms of the fields, the different terms of Eq.~\eqref{eq:phif} are given in the laboratory frame by
\begin{equation}
 \phi_\mathrm{L}=\mathbf{J}_\mathrm{f}\cdot\mathbf{E},
\end{equation}
\begin{equation}
 \phi_\mathrm{A}=\frac{1}{2}\Big(\mathbf{E}\cdot\frac{\partial\mathbf{P}}{\partial t}-\frac{\partial\mathbf{E}}{\partial t}\cdot\mathbf{P}+\mathbf{B}\cdot\frac{\partial\mathbf{M}}{\partial t}-\frac{\partial\mathbf{B}}{\partial t}\cdot\mathbf{M}\Big),
\end{equation}
\begin{equation}
 \mathbf{f}_\mathrm{L}=\rho_\mathrm{f}\mathbf{E}+\mathbf{J}_\mathrm{f}\times\mathbf{B},
\end{equation}
\begin{equation}
 \mathbf{f}_\mathrm{A}=\frac{\partial}{\partial t}\Big(\mathbf{P}\times\mathbf{B}-\frac{\mathbf{M}\times\mathbf{E}}{c^2}\Big)
 +\frac{1}{2}\nabla\times\boldsymbol{\tau},
 \label{eq:fA}
\end{equation}
% \begin{equation}
%  \mathbf{f}_\mathrm{A}=\frac{\partial}{\partial t}(\mathbf{G}_\mathrm{M}-\mathbf{G}_\mathrm{A})
%  +\frac{1}{2}\nabla\times(\mathbf{P}\times\mathbf{E}+\mathbf{M}\times\mathbf{B}),
% \end{equation}
% \begin{equation}
%  \mathbf{f}_{\boldsymbol{\tau}}=\frac{1}{2}\nabla\times(\mathbf{P}\times\mathbf{E}+\mathbf{M}\times\mathbf{B})
% \end{equation}
\begin{equation}
 \mathbf{f}_\mathrm{int}=\frac{1}{2}\Big[\mathbf{P}\cdot(\nabla\mathbf{E})-\mathbf{E}\cdot(\nabla\mathbf{P})+\mathbf{M}\cdot(\nabla\mathbf{B})-\mathbf{B}\cdot(\nabla\mathbf{M})\Big],
\end{equation}
\begin{equation}
 \mathbf{f}_\mathrm{ost}=\frac{1}{2}\nabla(\mathbf{P}\cdot\mathbf{E}+\mathbf{M}\cdot\mathbf{B}).
\end{equation}
% \begin{equation}
%  \mathbf{f}_\mathrm{mech}=-\nabla p_\mathrm{mech}.
% \end{equation}
Here the gradient of a vector is defined as $\nabla\mathbf{E}=\partial_x\mathbf{E}\otimes\hat{\mathbf{e}}_x+\partial_y\mathbf{E}\otimes\hat{\mathbf{e}}_y+\partial_z\mathbf{E}\otimes\hat{\mathbf{e}}_z$. The second term of $f_\mathrm{A}$ in Eq.~\eqref{eq:fA}, equal to $f_\mathbf{S}=\frac{1}{2}\nabla\times\boldsymbol{\tau}$, is determined by the spin-torque density in Eq.~\eqref{eq:torquedensity5}. This term is analogous to the spin-momentum contribution to the total momentum density of the electromagnetic field, when the total momentum density of the electromagnetic field in Eq.~\eqref{eq:Gem} is written as $\mathbf{G}_\mathrm{em}=\boldsymbol{\rho}_{\mathbf{p},\mathrm{em}}+\frac{1}{2}\nabla\times\boldsymbol{\rho}_{\mathbf{S},\mathrm{em}}$, where $\boldsymbol{\rho}_{\mathbf{p},\mathrm{em}}$ is the orbital momentum density. This splitting of the total momentum density of light into the orbital- and spin-momentum densities has been studied in previous literature \cite{Berry2009,Ohanian1986,Mita2000,Wittig2009,Bliokh2010,Bliokh2013b,Bliokh2017b}. The relations above show that the torque and force density dynamics are intimately related to each other.


\begin{thebibliography}{10}
\newcommand{\enquote}[1]{``#1''}

\bibitem{Beth1936}
R.~A. Beth, \enquote{Mechanical detection and measurement of the angular
  momentum of light,} \emph{Phys. Rev.} \textbf{50}, 115 (1936).

\bibitem{Allen2002}
L.~Allen and M.~J. Padgett, \enquote{Response to question \#79. does a plane
  wave carry spin angular momentum?} \emph{Am. J. Phys.} \textbf{70}, 567
  (2002).

\bibitem{Khrapko2001}
R.~I. Khrapko, \enquote{Does plane wave not carry a spin?} \emph{Am. J. Phys.}
  \textbf{69}, 405 (2001).

\bibitem{Landau1982}
V.~B. Berestetskii, E.~M. Lifshitz, and L.~P. Pitaevskii, \emph{Quantum
  Electrodynamics}, Pergamon, Oxford (1982).

\bibitem{Sakurai1967}
J.~J. Sakurai, \emph{Advanced Quantum Mechanics}, Addison-Wesley, Reading, MA
  (1967).

\bibitem{Dirac1958}
P.~A.~M. Dirac, \emph{The Principles of Quantum Mechanics}, Oxford University
  Press, Oxford (1958).

\bibitem{Messiah1961}
A.~Messiah, \emph{Quantum Mechanics}, Interscience, New York (1961).

\bibitem{Alpeggiani2018}
F.~Alpeggiani, K.~Y. Bliokh, F.~Nori, and L.~Kuipers, \enquote{Electromagnetic
  helicity in complex media,} \emph{Phys. Rev. Lett.} \textbf{120}, 243605
  (2018).

\bibitem{Bliokh2017b}
K.~Y. Bliokh, A.~Y. Bekshaev, and F.~Nori, \enquote{Optical momentum and
  angular momentum in complex media: from the {A}braham-{M}inkowski debate to
  unusual properties of surface plasmon-polaritons,} \emph{New J. Phys.}
  \textbf{19}, 123014 (2017).

\bibitem{Cameron2012}
R.~P. Cameron, S.~M. Barnett, and A.~M. Yao, \enquote{Optical helicity, optical
  spin and related quantities in electromagnetic theory,} \emph{New J. Phys.}
  \textbf{14}, 053050 (2012).

\bibitem{Jackson1999}
J.~D. Jackson, \emph{Classical Electrodynamics}, Wiley, New York (1999).

\bibitem{Bliokh2013b}
K.~Y. Bliokh, A.~Y. Bekshaev, and F.~Nori, \enquote{Dual electromagnetism:
  helicity, spin, momentum and angular momentum,} \emph{New J. Phys.}
  \textbf{15}, 033026 (2013).

\bibitem{Partanen2018a}
M.~Partanen and J.~Tulkki, \enquote{Mass-polariton theory of sharing the total
  angular momentum of light between the field and matter,} \emph{Phys. Rev. A}
  \textbf{98}, 033813 (2018).

\bibitem{Berry2009}
M.~V. Berry, \enquote{Optical currents,} \emph{J. Opt. A} \textbf{11}, 094001
  (2009).

\bibitem{Ohanian1986}
H.~C. Ohanian, \enquote{What is spin?} \emph{Am. J. Phys.} \textbf{54}, 500
  (1986).

\bibitem{Mita2000}
K.~Mita, \enquote{Virtual probability current associated with the spin,}
  \emph{Am. J. Phys.} \textbf{68}, 259 (2000).

\bibitem{Wittig2009}
C.~Wittig, \enquote{Photon and electron spins,} \emph{J. Phys. Chem. A}
  \textbf{113}, 15320 (2009).

\bibitem{Bliokh2010}
K.~Y. Bliokh, M.~A. Alonso, E.~A. Ostrovskaya, and A.~Aiello, \enquote{Angular
  momenta and spin-orbit interaction of nonparaxial light in free space,}
  \emph{Phys. Rev. A} \textbf{82}, 063825 (2010).

\bibitem{Partanen2023a}
M.~Partanen, B.~Anghinoni, N.~G.~C. Astrath, and J.~Tulkki,
  \enquote{Time-dependent theory of optical electro- and magnetostriction,}
  \emph{Phys. Rev. A} \textbf{107}, 023525 (2023).

\bibitem{Anghinoni2022}
B.~Anghinoni, G.~Flizikowski, L.~Malacarne, M.~Partanen, S.~Bialkowski, and
  N.~Astrath, \enquote{On the formulations of the electromagnetic
  stress–energy tensor,} \emph{Ann. Phys.} \textbf{443}, 169004 (2022).

\bibitem{Anghinoni2023}
B.~Anghinoni, M.~Partanen, and N.~Astrath, \enquote{The microscopic {A}mp\'ere
  formulation for the electromagnetic force density in linear dielectrics,}
  \emph{Eur. Phys. J. Plus} \textbf{138}, 1034 (2023).

\bibitem{Penfield1967}
P.~Penfield and H.~A. Haus, \emph{Electrodynamics of Moving Media}, MIT Press,
  Cambridge, MA (1967).

\bibitem{Partanen2021b}
M.~Partanen and J.~Tulkki, \enquote{Covariant theory of light in a dispersive
  medium,} \emph{Phys. Rev. A} \textbf{104}, 023510 (2021).

\bibitem{Kemp2017}
B.~A. Kemp and C.~J. Sheppard, \enquote{Electromagnetic and material
  contributions to stress, energy, and momentum in metamaterials,} \emph{AEM}
  \textbf{6}, 11 (2017).

\bibitem{Obukhov2008}
Y.~Obukhov, \enquote{Electromagnetic energy and momentum in moving media,}
  \emph{Ann. Phys.} \textbf{17}, 830.

\bibitem{Partanen2022b}
M.~Partanen and J.~Tulkki, \enquote{Time-dependent optical force theory for
  optomechanics of dispersive 3{D} photonic materials and devices,} \emph{Opt.
  Express} \textbf{30}, 28577 (2022).

\end{thebibliography}
\end{document}